\documentclass[aps,pra,twocolumn,amsmath,amssymb,superscriptaddress,floatfix,showpacs]{revtex4}
\pdfoutput=1
\usepackage{graphicx}
\usepackage{dcolumn}
\usepackage{bm}
\usepackage{stmaryrd}
\usepackage{latexsym}
\usepackage{amsfonts}

\begin{document}

\title{Many-body Landau-Zener tunneling in the Bose-Hubbard model}

\author{Andrea Tomadin}
\affiliation{%
Scuola Normale Superiore,
Piazza dei Cavalieri 7, I-56126 Pisa
}
\author{Riccardo Mannella}
\affiliation{%
Dipartimento di Fisica ``Enrico Fermi'',
Universit\`a degli Studi di Pisa, Largo Pontecorvo 3, I-56127 Pisa
}
\author{Sandro Wimberger}
\email{S.Wimberger@thphys.uni-heidelberg.de}
\affiliation{%
Institut f\"ur Theoretische Physik, Universit\"at Heidelberg,
Philosophenweg 19, D-69120 Heidelberg
}

\date{\today}

\begin{abstract}
We study a model for ultracold, spinless atoms in quasi-one dimensional optical lattices and subjected to a tunable tilting force.
Statistical tests are employed to {\em quantitatively} characterize the spectrum of the Floquet-Bloch 
operator of the system along the transition from the regular to the quantum chaotic regime.
Moreover, we perturbatively include the coupling of the one-band model to the second energy band.
This allows us to study the Landau-Zener interband tunneling within a truly many-body description of ultracold atoms.
The distributions of the computed tunneling rates provide an independent and experimentally accessible signature of the regular-chaotic transition.
\end{abstract}


%\keywords{}

\maketitle

\section{\label{sec:Intro} Introduction}

Bose-Einstein condensates loaded into optical lattices, which perfectly realize spatially periodic potentials, represent an exciting field of research in the sense that many simplified toy models of condensed matter physics can be studied in an exceptionally clean manner \cite{Group1}. 
This is achieved by modern means of atom and quantum optics that allow the experimentalist an unprecedented control of initial conditions in coordinate and momentum space and also of the desired dynamics \cite{Group1,SR2006}.

A paradigm of quantum transport on the microscopic scale is the Wannier-Stark (WS) problem, where particles move in a tilted, but otherwise spatially periodic potential. 
The famous Bloch oscillations and related phenomena, such as interband tunneling, were observed in experiments with quasi-particles in superlattices \cite{Group2}, with light in optical nonlinear media \cite{Group3}, and in great detail also with ultracold atoms moving in optical lattices \cite{Group1,Group4,Group5, RMFOMI2004, SZLWCMA2007}.
All experimental studies based on the latter realization were performed in a regime where atom-atom interactions are either negligible \cite{Group4} or they reduce to a perturbative mean-field effect \cite{Group5,RMFOMI2004,SZLWCMA2007}.

The regime of strong correlations in the WS system, in which interactions cannot be reduced to a mean-field model or even dominate the evolution has been addressed only theoretically up to now \cite{BK2003,KB2003,Kolovsky2003,PMKB2006}. 
State-of-the-art experiments are, however, capable of getting into a regime of filling factors (i.e. of atoms per lattice site) of the order one, where interaction-induced correlations are crucial \cite{Group6, BDZ2007}.

Motivated by the experimental progress, we extend previous studies of the asymmetric triple-well \cite{Group9} and of the WS problem \cite{BK2003,KB2003,Kolovsky2003}.
In the present work we give a more comprehensive and quantitative account of our findings briefly reported in \cite{TMW2007}.
In Section~\ref{sec:oneBand} we introduce our two-bands Bose-Hubbard (BH) model and focus on the dynamics within the first band of the optical lattice.
In contrast to the vast literature which focuses on regimes around Mott-Insulator like phase transitions in the absence of an additional Stark force, see e.g. \cite{FWGF1989,SSG2002,SS2005,Group7,Group8, Cederbaum1}, we concentrate on the BH model in the superfluid realm and in the presence of a static tilt.
We characterize the transition between the regular and the chaotic realm of the quantum spectra by a quantitative and systematical analysis based on statistical tests. 
In Section~\ref{sec:twoBands} we perturbatively include the decay to the second band via Landau-Zener like tunneling processes.
The quantum spectrum of the latter, non-unitary problem is analyzed in Section~\ref{sec:Results} and found to essentially reproduce the properties of the purely one-band approximation down to rather small lattice depths. 
The resulting decay rates for the interband tunneling strongly depend on the many-particle nature of the problem and are found to correlate with the transition to the quantum chaotic regime.
As a consequence, signatures of many-body quantum chaos are predicted to be accessible to experiments with ultracold atoms over a broad range of parameters, in both ``horizontal'' transport along the lattice and in interband ``vertical'' transport.
Our results are finally summarized in Section \ref{sec:Conclusion}.

\section{\label{sec:oneBand} Spectral Analysis of the one-band Bose-Hubbard model}

We briefly review the general Hamiltonian for a system of spinless, interacting atoms in a quasi-one dimensional optical lattice subjected to a tunable tilting force $F$. 
We start out with the purely periodic problem, $F=0$, in an optical potential of spacing $a$, recoil momentum $k_{L}=\pi/a$ and typical kinetic energy $E_{R}=k_{L}/2m$. 
The optical potential and the kinetic energy form the single-particle Bloch Hamiltonian:
\begin{equation}\label{eq:oneBand1}
H_{\rm one }(x)=- \frac{\triangle}{2m} - V\,\cos\left ( \frac{2\pi x}{a} \right ).
\end{equation}
The eigenfunctions are the Bloch waves $\psi$, labeled by the quasimomentum $k$ and the band index $\alpha$ \cite{GKK2002}, with dispersion law $E_{k,\alpha}$.
In the Appendix~\ref{sec:Wannier} we explain how to use the single-particle solutions $\psi_{k,\alpha}$ to build a set of localized orbitals $\chi_{\ell,\alpha}$ called Wannier Functions (WF). 
In the limit of deep lattices, the orbital $\chi_{\ell,\alpha}$ goes to the wave function of the $\alpha-$th excited level for an harmonic potential centered on the $\ell-$lattice site.
We use the WF to expand the field of the ultracold bosons:
\begin{equation}\label{eq:oneBand2}
\hat{\phi }(x)=\sum_{\alpha}\sum_{\ell}\chi_{\ell,\alpha}(x)\hat{a}_{\ell\alpha}.
\end{equation}
Then we introduce a Stark force $F$ that tilts the optical potential and a zero-range interaction between the atoms, parametrized by the scattering length $a_{S}$. 
In a quasi-one dimensional optical lattice -- as realizable in experiments \cite{BDZ2007} -- the scattering length is derived from the true three-dimensional scattering length
via a renormalization that accounts for the transverse confinement of the atomic wave functions \cite{BDZ2007} and the physical
dimension is then $[a_{S}]=L^{-1}$. 
The Hamiltonian in the second quantization is written  as:
\begin{eqnarray}\label{eq:oneBand4}
\hat{H}  
& = &  \int \hat{\phi}^{\dag}(x)  \left [ H_{\rm one}(x)+Fx \right ] \hat{\phi }(x) {\rm d}x \nonumber \\
& + & \frac{1}{2}\frac{4\pi a_{S} }{m} \int  \hat{\phi }^{\dag}(x) \hat{\phi }^{\dag}(x) \hat{\phi }(x) \hat{\phi}(x)  {\rm d}x. 
\end{eqnarray}
Substituting the expansion Eq.~\eqref{eq:oneBand2} into Eq.~\eqref{eq:oneBand4} we obtain the Hamiltonian in terms of the creation and annihilation operators $\hat{a}_{\ell,\alpha}^{\dag}$, $\hat{a}_{\ell,\alpha}$, for a particle in the $\ell-$th site  and the $\alpha$-th energy band of the lattice.
The number operator is $\hat{n}_{\ell,\alpha}=\hat{a}_{\ell,\alpha}^{\dag}\hat{a}_{\ell,\alpha}$.
We restrict the analysis to the first two bands, which can be addressed by experiments \cite{MFWB2007} and that can be handled numerically without great difficulties.
The coefficients of the Hamiltonian are given by integrals involving the WF: the exact computation of the WF outlined in Appendix~\ref{sec:Wannier} motivates the selection of the operators that are most relevant for $ V \gtrsim E_{R}$.
We are left with the on-site energy 
$\hat{a}_{\ell,\alpha}^{\dag}\hat{a}_{\ell,\alpha}$, the kinetic energy
$\hat{a}_{\ell+1,\alpha}^{\dag}\hat{a}_{\ell,\alpha}$,
and the on-site interaction between atoms in the same band 
$\hat{n}_{\ell,\alpha}(\hat{n}_{\ell,\alpha} -1)$ \cite{JBCGZ1998}, for $\alpha\in\{1,2\}$.
Moreover we have on-site interaction between atoms in different bands $\hat{n}_{\ell,1}\hat{n}_{\ell,2}$ and two transition operators $\hat{a}_{\ell,2}^{\dag}\hat{a}_{\ell,1}$,
$\hat{a}_{\ell,1}^{\dag}\hat{a}_{\ell,1}^{\dag}\hat{a}_{\ell,2}\hat{a}_{\ell,2}$ that are the subject of a detailed analysis in Section~\ref{sec:twoBands}.

The dimension $D_{\rm H}$ of the Hilbert space spanned by the Fock states $|\vec{n}\rangle$ (defined in Appendix~\ref{sec:FloquetBloch}), for a system of $N$ bosons distributed over $L$ lattice sites, occupying up to the $2$nd band of the periodic potential, is given by the combinatorial formula  $D_{\rm H } = (N+2L-1)!/N!(2L-1)!$.
The typical number of lattice sites in experiments is $L\lesssim 100$ and the filling-factor $N/L$ is of order unity \cite{Group6, BDZ2007}, such that the exponential increase of $D_{\rm H}$ with the system size limits any exact numerical approach to smaller systems, where we impose the cyclic boundary conditions $\hat{a}_{L,\alpha }=\hat{a}_{0,\alpha}$.
The implementation of these conditions requires the system to be translationally invariant.
The Stark potential $Fx$, however, spoils the periodicity of the Bloch Hamiltonian Eq.~\eqref{eq:oneBand1} and produces localized WS eigenstates instead of traveling Bloch waves \cite{GKK2002}.
We follow \cite{KB2003} and proceed to eliminate the Stark potential from the Hamiltonian by changing to the Interaction Representation (IR) with respect to 
$\hat{H}_{\rm S}=F\int\hat{\phi}^{\dag}x\hat{\phi}\,{\rm d}x=aF\;\sum_{\ell=1}^{L}\sum_{\alpha}\ell \hat{n}_{\ell,\alpha }$.
The Hamiltonian in the IR, $\hat{H}(t)=e^{-i\hat{H}_{\rm S}t}\hat{H}e^{+i\hat{H}_{\rm S}t}$, is time-dependent, and the problem becomes conceptually more complicated.

We rescale the energies by $E_{R}$, the lengths by $a$, the momenta by $k_{L}$ and we make the substitutions $F\leftarrow FE_{R}/a$, $\chi\leftarrow \chi/\sqrt{a}$.
The on-site energies $\varepsilon_{\alpha}$ and the hopping amplitudes $J_{\alpha}$ are given in Eq.~\eqref{eq:Wannier3}.
The interaction coefficients $W_{\alpha}$, $W_{\times}$, are proportional to the coupling constant $W = 4\pi a_{S}/am E_{R}$ and are given in Eq.~\eqref{eq:Wannier4}.
The ``dipole'' coefficient $d_{F}$ is given in Eq.~\eqref{eq:Wannier5}.
The Hamiltonian of Eq.~\eqref{eq:oneBand4}, restricted to the first two bands of the lattice,  finally reads in the IR:
\begin{widetext}
\begin{eqnarray}\label{eq:oneBand3}
\hat{H}(t)  
& = &  \sum_{\ell=1}^{L} \left \lbrace
\varepsilon_{1}\hat{n}_{\ell,1} 
	-\frac{1}{2} J_{1}
		e^{iFt } \hat{a}_{\ell+1,1}^{\dag}\hat{a}_{\ell,1}
		+ \; \mbox{H.c.}
	+\frac{1}{2}W_{1} \hat{n}_{\ell,1}(\hat{n}_{\ell,1}-1) 
	+(1\rightarrow 2)
	\right. \nonumber \\
& + & \left. 2W_{\times}  \hat{n}_{\ell,1}\hat{n}_{\ell,2}
	+F d_{F} ( \hat{a}_{\ell,2}^{\dag}\hat{a}_{\ell,1}  + \; \mbox{H.c.}   ) 
	+ \frac{1}{2}W_{\times}  
		( \hat{a}_{\ell,1}^{\dag}\hat{a}_{\ell,1}^{\dag}\hat{a}_{\ell,2}\hat{a}_{\ell,2} + \; \mbox{H.c.}  )
	\right \rbrace .
\end{eqnarray}
\end{widetext}
In the IR the Hamiltonian is again symmetric for discrete translations in space and it has lost the time independence but it is periodic with the Bloch period $T_{\rm B}=2\pi/F$.
\begin{figure}
	\includegraphics[width=\columnwidth]{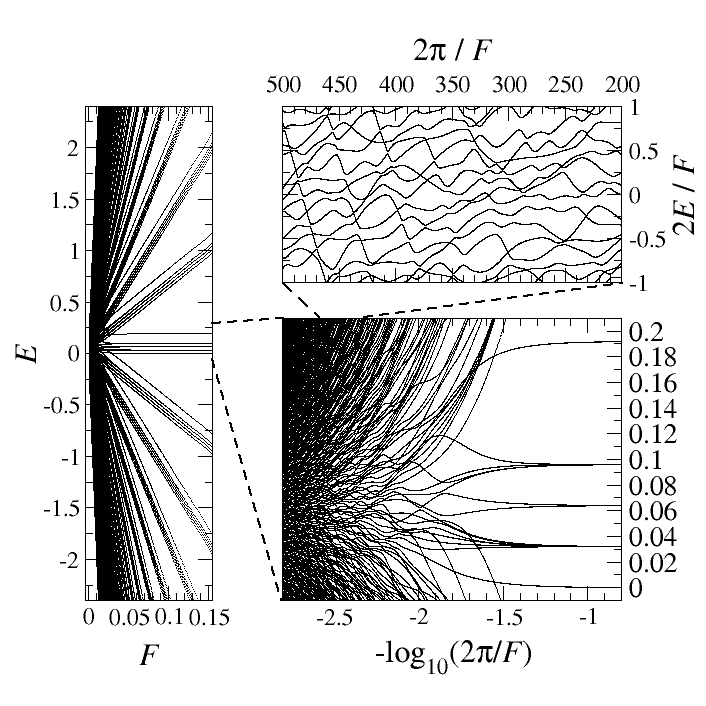} %
	\caption{\label{fig:oneBand1}
	The quasienergy levels $E$ of the FB operator for a system with $N=4$, $L=5$, in the subspace with $\kappa=0$.
	The WS ladder is seen on the left panel, the perturbative splitting of the first rung is magnified in the lower panel.
	In the upper panel a single linear function of $1/F$ is added to $2E/F$ to eliminate an overall winding trend, and allow a better visibility of the avoided crossings.
	The parameters of the Hamiltoniana are $J_{1}=0.038$, $W_{1}=0.032$, as used also in the subsequent Figures.
	}
\end{figure}
We assume that the initial state, for $F=0$, is superfluid, characterized by the delocalization of the atoms over the entire lattice.
The critical conditions on the parameters, that enforce the superfluid phase in the present context \cite{FWGF1989} is $W_{1} / J_{1} \leq 5.8$.
Following \cite{BK2003}, in the present Section we set the lattice depth $V\simeq 5$, which gives $J_{1}\simeq 0.038$ and the interaction coefficient $W\simeq 0.016$, with $W_{1}\simeq 0.032$.

The object of the subsequent study is the evolution operator up to the Bloch period $\hat{\cal U}_{\rm FB} = \widehat{\exp}(-\int_{0}^{{T}_{\rm B}}i\hat{H}(t) {\rm d}t)$, called Floquet-Bloch (FB) operator \cite{KB2003}.
The results presented in the following confirm and extend the results of \cite{KB2003}.
The discrete translational symmetry of the Hamiltonian entails that the FB operator is a block-diagonal matrix in the basis of Eq.~\eqref{eq:FloquetBloch2}, labeled by a many-body quasimomentum $\kappa$.
The dynamics of the atoms in the lattice is complex because many vectors take part in the time evolution of an arbitrary initial state.
The strong mixing of the basis vectors in time means that the evolution of a state is not bound to a small subspace of the total Hilbert space (contrary to \cite{SSG2002}), but, after a Bloch period, the initial state spreads over the entire Hilbert subspace with definite quasimomentum (e.g. $\kappa=0$).
This is evidenced by the dependence on $F$ of the quasienergies $E_{j}$, obtained from the eigenvalues ${\rm exp}(-iE_{j}T_{\rm B})$ of the FB operator.

In the Fig.~\ref{fig:oneBand1} we show the quasienergy spectrum as the ``control parameter'' $F$ is varied.
In the limit $F\rightarrow 0$, we recover the standard BH model (the analysis of the FB spectrum is here, however, not useful since the Bloch period tends to infinity).
In the regime where the atomic interactions are negligible with respect to the lattice potential, the single particle Bloch picture is adequate and the spectrum is simply a finite band.
For $F\gtrsim 0.1$ the single-particle WS ladder is found, i.e. a fan of energy levels $E_{m}(F)\simeq 2\pi m F$, $m$ integer.
Since the interactions are non-zero, the levels are split up and the first order perturbative effect on the ladder was computed in \cite{Kolovsky2003}.
The central WS rung is split into levels which are proportional to the interaction energies of many atoms in a site, $W_{1}n(n-1),\; n=1,\dots,4$.
Since the level splittings have the common factor $W_{1}$, a collapse and revival of quasimomentum oscillations (the Fock space version of the single-particle Bloch Oscillations) was predicted for this parameter range in \cite{Kolovsky2003}.
On the contrary, an irreversible decay of the quasimomentum oscillations was found in the range $F \simeq W_{1}$ \cite{BK2003}, where avoided crossings dominate the spectrum.
These are directly, experimentally observable consequences of the complex level structure presented in Fig.~\ref{fig:oneBand1}.
\begin{figure} 
  \centering
  \includegraphics[width=\columnwidth]{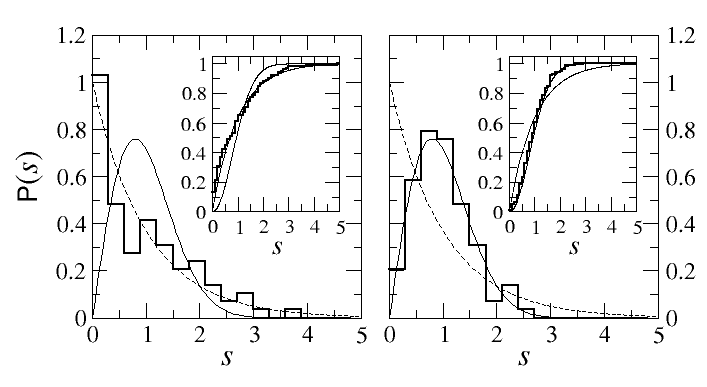}
  \caption{\label{fig:oneBand2}
	The distribution $\mathsf{P}(s)$ and the $\mathsf{CDF}$ (insets) for the quasienergy spacings (stairs), the WD (solid) and the Poisson distribution (dashed).
	The parameters are $N=5$, $L=8$, $\log_{10}(2\pi/F)\simeq 2.0$ (left, regular) and $2.4$ (right, chaotic).
	}
\end{figure}
The presence of avoided crossings means that a strong mixing of the Fock states is necessary to build the eigenstates of the system, and no set of quantum numbers can be assigned to individual levels as $F$ varies.
In the region of parameters where the energy scales of the system are comparable in magnitude, we can characterize the spectrum in terms of the statistical distribution of the quasienergies spacings and by statistical measures for the eigenfunctions.
The latter have been intensively studied in \cite{Group20}.
In the following, we concentrate on the statistical behaviour of the quasienergies, which is closely linked to the behaviour of the open system studied in Section~\ref{sec:twoBands}.

The probability $\mathsf{P}(s){\rm d}s$ that the magnitude of a given interval spacing $s_{j}=\Delta E_{j} / \langle \Delta E_{j} \rangle_j$ is in $[s,s+{\rm d}s]$ is given by the Poisson distribution $\mathsf{P}(s) = \exp(-s)$ \cite{Haake2001} for an uncorrelated spectrum in the regular regime.
Strongly correlated quantum spectra, corresponding to the chaotic regime in our many-particle model, are well modeled by the Wigner-Dyson (WD) distribution for a Circular Orhogonal Ensemble of random matrices \cite{Haake2001}: $\mathsf{P}(s) = \pi\,s\, {\rm exp}(-\pi\,s^{2}\,/4)/2$. 
In Fig.~\ref{fig:oneBand2} the probability distribution and the Cumulative Distribution Function ($\mathsf{CDF}$) of the quasienergy spacings are shown for two paradigmatic values of $F$.
The presence of avoided crossings in the chaotic case $\log_{10}(2\pi/F)\simeq 2.4$ shows up as a depletion of small quasienergy spacings, and the probability to find a level crossing vanishes.

\begin{figure}
	\centering
	\includegraphics[width=\columnwidth]{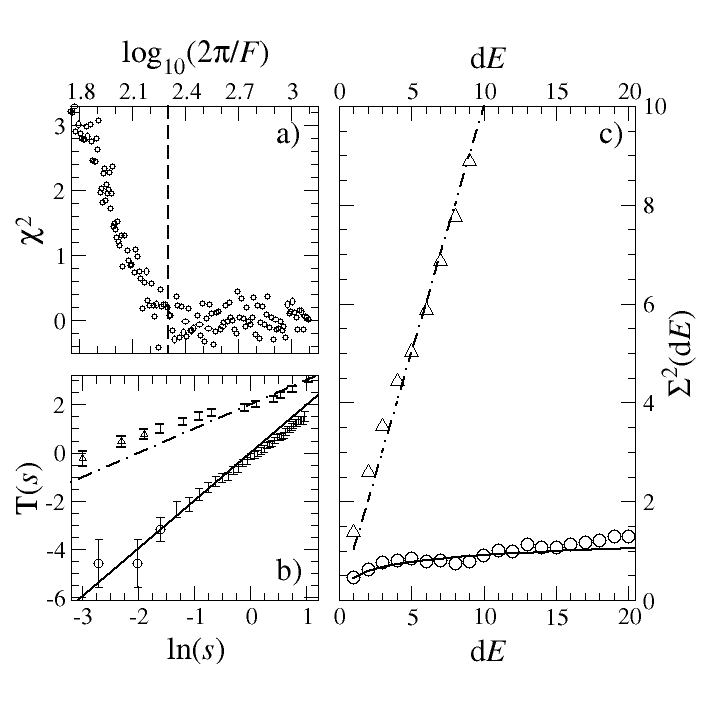}
	\caption{\label{fig:oneBand3}
	Statistical tests applied to the system of Fig.~\ref{fig:oneBand2}.
	(a) $\chi^{2}-$like test of Eq.~\eqref{eq:oneBand6}.
	(b) $T$ test and (c) variance of number of levels (with mean spacing normalized to 1), for the chaotic (circles) and the regular (pyramids) case. 
	In both panels the solid and dot-dashed lines are the theoretical predictions for the WD and Poisson statistics of energy levels, respectively. 
	}	 
\end{figure}

We improved the statistical description of the quasienergy spectrum with further analyses, shown in Fig.~\ref{fig:oneBand3}. 
In the panel (a) we quantify the convergence of the quasienergy spacings distribution to the WD profile, thus filling the gap between the two pictures of Fig.~\ref{fig:oneBand2}.
We computed the FB operator for several values of $F$ and confronted each spectrum with the WD distribution using a modified $\chi^{2}$ test, computed as follows. 
Each sequence of levels spacings was algorithmically binned to leave $5\leq O_{b} \leq 10$ ``observed'' spacings in each bin $b=1,\dots,N_{b}$ \cite{NumericalRecipes}.
The ``expected'' values $E_{b}$ are the integrals of the theoretical distributions over the bins and the sum $Q=\sum_{b}\left ( O_{b}-E_{b} \right )^{2}/E_{b}$ was calculated.
The values of $Q$ are distributed according to a $\chi^{2}$ distribution with $N_{b}-1$ degrees of freedom and mean $N_{b}-1$.
The renormalized variable
\begin{equation}\label{eq:oneBand6}
\chi^{2}=\log_{10}[Q/(N_{b}-1))]
\end{equation}
is thus appropriate to compare several data sets, each binned optimally and independently, 
For $F \lesssim 0.025$, $\chi^{2}$ is in the range $[-0.5,0.5]$ (in the bulk of the original $\chi^{2}$ distributions before the transformation of Eq.~\eqref{eq:oneBand6} was performed), and the correspondent distribution of the spacings is well described by a WD profile.
As the external force increases and the parameter $2\pi/F$ diminishes, the larger values of $\chi^{2}$ 
(lying in the tails of the original $\chi^{2}$ distributions) indicate that the spectrum is not well characterized by a WD distribution.
The condition for the regular-chaotic transition can be directly read off from the quantitative statistical test results and corresponds, e.g. to $\log_{10}(2\pi/F) \simeq 2.3$.

We found that the profile of the quasienergy spacings distribution changes smoothly, but we used two statistical tests introduced in \cite{PR1993} (eq. (27) and (29) therein) to show that the Poisson and WD statistics are clearly identified at the borders of the transition.
Fig.~\ref{fig:oneBand3} (b) shows the explicit results for the $T$ function of \cite{PR1993} whose predicted linear scaling $T\sim \ln s$ for the Poisson and $T\sim 2 \ln s$ for the WD expectation is confirmed. 
In Fig.~\ref{fig:oneBand3} (c) we show the variance $\Sigma^{2}({\rm d}E)$ \cite{Mehta1991} of the number of levels $N(E,E+{\rm d}E)$ found in a finite energy interval ${\rm d}E$:
\begin{equation}\label{eq:oneBand7}
\Sigma^{2}({\rm d}E) \equiv \langle [N(E,E+{\rm d}E)-\bar{N}{\rm d}E]^2 \rangle_{E},
\end{equation} 
where the average is taken over all the energies and we rescaled the spectrum so that the average number of levels
per unit of energy $\bar{N}$ is equal to unity.
A linear and logarithmic scaling is predicted for a Poisson and WD-like spectra, respectively. The logarithmic behaviour clearly prevails over all energy ranges for the chaotic spectrum, only apart from oscillations which are indeed typical for samples of finite size (see \cite{Mehta1991} for details).

\section{\label{sec:twoBands} Interband coupling in the many-body regime}

We develop a perturbative analysis of the two-bands system of Eq.~\eqref{eq:oneBand3}.
We consider the many-body dynamics within the ground band and the perturbative action of the operators that couple the Fock states of the ground band to states in the second band.
As a consequence of the interband terms each Fock state $|\vec{n}\rangle$ (see Appendix~\ref{sec:FloquetBloch}) suffers an energy shift $\delta E(\vec{n}) -i \Gamma_{\rm F}(\vec{n})/2$, where $\Gamma_{\rm F}(\vec{n})$ is its decay width.
The Hamiltonian is modified accordingly and becomes a non-Hermitian effective Hamiltonian for the ground band, that yields a non unitary FB operator.
In the following, we compute the set of decay widths.
In the next Section we then study the spectrum of this new FB operator.

We first define a basis of unperturbed states.
We choose to neglect the hopping in the lower band, where the WF are more strongly localized, so an unperturbed state projected on the Hilbert space of $N$ particles in the ground band is a Fock state $|\vec{n};N\rangle$.
In the second band we neglect the interactions, since in the perturbative approach only a few  particles (one or two in the following) populate the excited levels. 
So an unperturbed state projected on the second band is the solution of the one-particle WS problem \cite{SSG2002,GKK2002}, i.e. a localized wave function centred at site $w$, written with the Bessel function ${\cal J}_{m}(x)$ as:
\begin{equation}\label{eq:twoBands3}
|w\rangle=\sum_{\ell=-\infty}^{+\infty}{\cal J}_{\ell-w}(-J_{2}/F)\hat{a}_{\ell,2}|{\rm vac}\rangle.
\end{equation}
We approximate the Hilbert space of the system as the tensor product of the spaces of the two bands and the entanglement between the ground and the excited particles is neglected.
Then an unperturbed state with one or two promoted particles is of the form $|\vec{n};N-1\rangle\otimes |w\rangle$ or $|\vec{n};N-2\rangle\otimes |w,w'\rangle$, respectively.
In the following, the occupation number $n_{\ell,1}$ for the $\ell-$th site in the ground band is written $n_{\ell}$.

The Hamiltonian Eq.~\eqref{eq:oneBand3} contains two mechanisms that promote particles to the second band.
The first is a single-particle effect, a consequence of the external force, proportional to the dipole coupling $d_{F}$ between the WF of different bands. 
The Hamiltonian of the perturbation is
\begin{equation}\label{eq:twoBands1}
\hat{H}_{1}=Fd_{F} \sum_{\ell }  ( \hat{a}_{\ell,2}^{\dag}\hat{a}_{\ell,1} + \hat{a}_{\ell,1}^{\dag}\hat{a}_{\ell,2} ).
\end{equation}
The second perturbation is a many-body effect, describing two particles of the first band that collide and transform their interaction energy into kinetic energy, entering the second band together:
\begin{equation}\label{eq:twoBands2}
\hat{H}_{2}=\frac{1}{2}W_{\times} \sum_{\ell=1}^{L} (\hat{a}_{\ell,2}^{\dag}\hat{a}_{\ell,2}^{\dag}\hat{a}_{\ell,1}\hat{a}_{\ell,1} + \;\mbox{H.c.}  ).
\end{equation}
The expectation value of $\hat{H}_{1}$, $\hat{H}_{2}$ on the unperturbed states, equal to the first-order energy shift $\delta E(\vec{n})$, is zero because the operators do not conserve the number of particles $n_{\alpha}$ within the bands.

Let us focus on $\hat{H}_{1}$ and compute its matrix element for the channel:
\begin{equation}\label{eq:twoBands4}
|\vec{n};N\rangle\otimes |{\rm vac}\rangle\rightarrow |\vec{n}';N-1\rangle\otimes |w\rangle,\;
n'_{h}=n_{h}-1.
\end{equation}
The decay width at first-order is given by the Fermi's Golden Rule and only the first term in Eq.~\eqref{eq:twoBands1} gives nonzero contribution for the channel of Eq.~\eqref{eq:twoBands4}.
Our result for the matrix element is:
\begin{eqnarray}\label{eq:twoBands5}
&  &\langle k | \langle \vec{n}' |  \sum_{\ell=1}^{L} ( \hat{a}_{\ell,2}^{\dag}\hat{a}_{\ell,1} ) |\vec{n}\rangle|{\rm vac}\rangle=   \\ 
&  & = \sum_{\ell=1}^{L} {\cal J}_{\ell-w}(-J_{2}/F)\;\delta(n'_{\ell},n_{\ell}-1)\;\sqrt{n_{\ell }} \;\prod_{m\neq \ell } \delta(n'_{m},n_{m}). \nonumber
\end{eqnarray}
The Kronecker $\delta(\cdot,\cdot)$ functions act as a selection rule for the Fock states that are coupled by the perturbation.
The tunneling mechanism does not include any income of energy from an external source, so the initial and final energies,
\begin{eqnarray}\label{eq:twoBands6}
E_{0}(\vec{n}) & = & \langle{\rm vac}| \langle \vec{n} | \hat{H }_{0} |\vec{n} \rangle|{\rm vac}\rangle,\nonumber \\
E_{0}(\vec{n}',w) & = & \langle w | \langle \vec{n}' | \hat{ H }_{0} |\vec{n}'\rangle |w\rangle,
\end{eqnarray}
must be equal as required by the Golden Rule.
The condition on the energy conservation is relaxed to account for the uncertainty $\Delta E(\vec{n})$ of the unperturbed energy levels of the initial and final states.
The energy uncertainty and the level density function $\rho(E,\vec{n})$ are derived from the perturbative action of the hopping operator of the first band that has been neglected so far.
We postpone the computation of these quantities to the end of the present Section.
The relaxed energy conservation rule selects from Eq.~\eqref{eq:twoBands5} the set $K$ of permitted decay channels $(h,w)$ parametrized by the two indices $h,w$ such that:
\begin{eqnarray}\label{eq:twoBands7}
& & E_{0}(\vec{n}',w)-E_{0}(\vec{n}) = \nonumber \\ 
& & =   \varepsilon_{2}-\varepsilon_{1}-F (h-w)-W_{1} ( n_{h}-1 ) \nonumber \\
& & \in\; \left [-\frac{\Delta E(\vec{n})+\Delta E(\vec{n}')}{2},\,\frac{\Delta E(\vec{n})+\Delta E(\vec{n}')}{2}   \right ].
\end{eqnarray}
The last equation means that the energy $\varepsilon_{2}-\varepsilon_{1}$ required to promote a particle to the second band  is supplied by the decrease of the repulsion energy (proportional to $W$) and by the work of the force (proportional to $F$) exerted on the promoted particle.
 
The total width $\Gamma_{1}(\vec{n})$ for the decay via the allowed channels $K$, is proportional to the square of the matrix element and to the level density $\rho(E,\vec{n})$ given below in Eq.~\eqref{eq:twoBands16}.
We arrive at
\begin{eqnarray}\label{eq:twoBands8}
& & \Gamma_{1}(\vec{n})=2\pi  F^{2} d_{F}^{2} \\
& & \times \sum_{(h,w)\;\in \;K} 
	\left \lbrace \left | {\cal J}_{h-w}(-J_{2}/F) \cdot \sqrt{n_{h}} \right |^{2}
	\frac{1}{\Delta E(\vec{n})\Delta E(\vec{n}')} \right \rbrace. \nonumber
\end{eqnarray}
The perturbation $\hat{H}_{2}$ is treated in a similar way, with the difference that two particles are promoted to the second band, and the position of the second Stark state $| w'\rangle$ is an additional degree of freedom for the transition. 
The decay channels are:
\begin{eqnarray}\label{eq:twoBands9}
& & |\vec{n},N\rangle \otimes |{\rm vac}\rangle\rightarrow |\vec{n}',N-2\rangle \otimes | w,w' \rangle, \nonumber \\
& & n'_{h}=n_{h}-2.
\end{eqnarray}
The approximate energy matching equation selects a set $K$ of possible decay channels, parameterized by the three site indices $(h,w,w')$:
\begin{eqnarray}\label{eq:twoBands10}
& & (h,w,w')\;\in \;K\;\mbox{ s.t. }\; E_{0}(\vec{n}',w,w')-E_{0}(\vec{n})= \nonumber \\
& & = 2 (\varepsilon_{2}-\varepsilon_{1}) -F (2h-w-w')-W_{1}( 2n_{h}-3 ) \nonumber \\
& & \in\; \left [-\frac{\Delta E(\vec{n})+\Delta E(\vec{n}')}{2},\,\frac{\Delta E(\vec{n})+\Delta E(\vec{n}')}{2}   \right ].
\end{eqnarray}
We state the result for the decay width:
\begin{eqnarray}\label{eq:twoBands11}
&  & \Gamma_{2}(\vec{n})=\frac{1}{2}\pi W_{ \times}^{2} \hspace{-0.4cm}
\sum_{(h,w,w')\;\in \;K} 
\left \lbrace 
\left | {\cal J}_{h-w}(-J_{2}/F) \right . 
\phantom{\frac{1}{\Delta E(\vec{n})\Delta E(\vec{n}')}} 
\right . \nonumber \\
&  &
\left . \times
\left . {\cal J}_{h-w' }(-J_{2}/F) \right |^{2}
\,4 n_{h} \left (  n_{h}-1   \right )
\frac{1}{\Delta E(\vec{n})\Delta E(\vec{n}')} \right \rbrace.
\end{eqnarray}
With respect to Eq.~\eqref{eq:twoBands8}, the additional degree of freedom $w'$ results in an extra summation extended over the (infinite) possible values of the difference $w-w'$. 
This follows from the possibility to conserve the energy even if a particle is pushed very far, if the other particle is pushed almost equally far in the opposite direction.
Since the decay width Eq.~\eqref{eq:twoBands11} depends on the product of two (rapidly decaying) Bessel functions we apply the truncation $|w-w'| \leq |J_{2}/F |$, to reduce the formula to a finite form.

Now we conclude the computation and derive the energy broadening $\Delta E(\vec{n})$ of the Fock states in the ground band necessary to implement the conditions of Eq.~\eqref{eq:twoBands7} and \eqref{eq:twoBands10}.
In the ground band, the unperturbed Hamiltonian consists only of the on-site interaction operator:
\begin{displaymath}\label{eq:twoBands12}
\hat{H }_{0} =  \frac{1}{2} W_{1}  \sum_{\ell } \hat{n}_{\ell,1}(\hat{n}_{\ell,1}-1).
\end{displaymath}
In the case of a single particle in a periodic potential the use of first order perturbation theory is wrong, as the second order of the perturbation theory diverges because of the exact degeneracy in energy of neighbouring sites, entailed by the translational symmetry of the lattice.
On the contrary, in the present case, the unperturbed Hamiltonian is just a rough approximation of the true Hamiltonian of the system and the remaining operators are supposed to remove the degeneracies, since the translational symmetry is broken by the external field in the WS picture.
The perturbation Hamiltonian is given by 
\begin{equation}\label{eq:twoBands13}
\hat{H }_{h} =  -\frac{1}{2}J_{1} \sum_{\ell } ( \hat{a}_{\ell+1,1}^{\dag}\hat{a}_{\ell,1} + \;\mbox{H.c.} ),
\end{equation} 
and its matrix elements between Fock states are
\begin{eqnarray}\label{eq:twoBands14}
& &\langle \vec{n}' | \hat{H }_{h} | \vec{n} \rangle = 
-\frac{1}{2} J_{1} \sum_{\ell=1}^{L}\, \sum_{\Delta \ell =\pm 1} \hspace{-0.3cm}
\prod_{ \begin{array}{c} \scriptstyle{m\neq \ell } \\ \scriptstyle{m\neq \ell+\Delta \ell } \end{array} } \nonumber \\
& & \sqrt{n_{\ell,1}}\,\sqrt{n_{\ell+\Delta \ell,1}+1}\, \delta(n'_{m,1},n_{m,1}) \times \nonumber \\
& & \times \delta(n'_{\ell,1},n_{\ell,1}-1) \,\delta (n'_{\ell+\Delta \ell,1},n_{\ell+\Delta l,1}+1).
\end{eqnarray}
\begin{figure}
  \centering
  \includegraphics[width=\columnwidth]{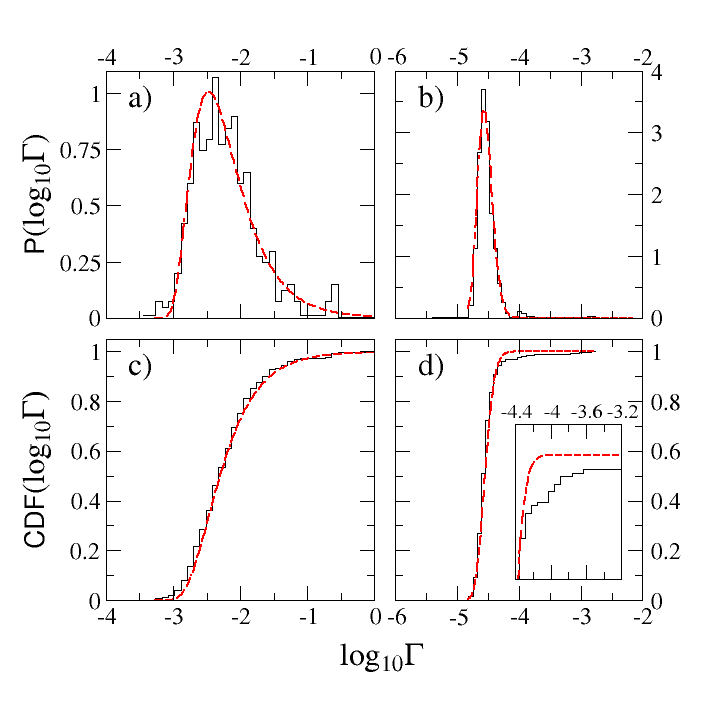}
	\caption{\label{fig:Results1}
	(a,b) the probability distribution $\mathsf{P}$ and (c,d) the $\mathsf{CDF}$  for the \emph{logarithm} of the decay widths $\Gamma$ in the regular regime [(a,c) with $\log_{10}(2\pi/F) \simeq 1.5$] and in the chaotic regime [(b,d) with $\log_{10}(2\pi/F) \simeq 2.1$].
	The size of the system is $N=8$, $L=7$.
	The dashed line is the fit with a lognormal distribution. The inset in panel (d) shows that the lognormal is not appropriate to fit the tails of the distributions in the chaotic regime.
	Here and in the following Figures, $V=1.5$, $\varepsilon_{2}-\varepsilon_{1}=2.63$, 
	$J_{1}=0.22$, $J_{2}=-1.0$, $W_{1}=0.2$, $W_{\times}=0.1$, $d_{F}=-0.2$.
	}
\end{figure}
Transitions are allowed between Fock states that differ for one boson in two adjacent holes $m$, $m+\Delta m$.
The transition channels are written as
$| \vec{n} \rangle \rightarrow |\vec{n}'\rangle$, with 
$\vec{n}'_{m,1}=\vec{n}_{m,1}-1$ and 
$\vec{n}'_{m+\Delta m,1}=\vec{n}_{m+\Delta m,1}+1$,
and must fulfil the condition of the energy conservation:
\begin{equation}\label{eq:twoBands15}
E_{0}(\vec{n}')-E_{0}(\vec{n}) = W_{1} \, ( n_{m+\Delta m,1} - n_{m,1} +1  ) \equiv 0.
\end{equation}
The total uncertainty is obtained by adding up the contributions from all the transitions and the 
summation over the allowed channels can be re-casted in a summation over the lattice sites.
Using the Golden Rule, each Fock state is attributed the following energy uncertainty:
\begin{eqnarray}
& & \Delta E(\vec{n}) = \frac{1}{2}\pi  J_{1}^{2} \; \sum_{\vec{n}'}\,\Delta E (\vec{n}\rightarrow \vec{n}' ) = \nonumber \\
& & = \frac{1}{2}\pi  J_{1}^{2}\; \sum_{\ell }\,\sum_{\Delta \ell=\pm 1} n_{\ell,1}^{2}\; \delta(n_{\ell+\Delta \ell,1}+1,n_{\ell,1}).
\end{eqnarray}
Finally, the level density $\rho(E,\vec{n})$ around the unperturbed energy $E_{0}(\vec{n})$ of a Fock states $| \vec{n} \rangle$ is approximated by a rectangular profile, of width $\Delta E(\vec{n})$ and area unity:
\begin{equation}\label{eq:twoBands16}
\rho(E,\vec{n}) =  \chi \left \lbrace |E-E_{0}(\vec{n})|\leq \Delta E(\vec{n}) / 2 \right \rbrace\; /\; \Delta E(\vec{n}).
\end{equation}

\section{\label{sec:Results} Results of the perturbative opening of the one-band system}

\begin{figure} 
  	\centering
  	\includegraphics[width=\columnwidth]{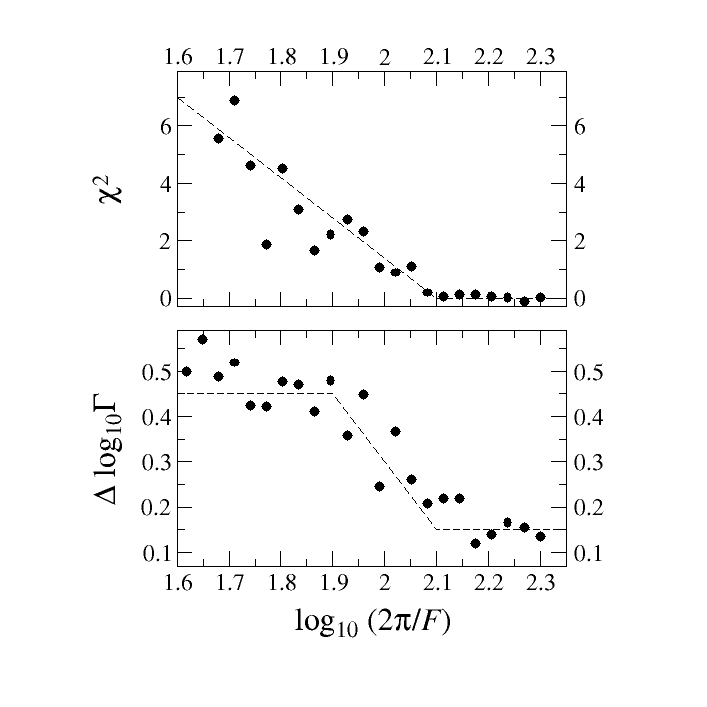}
	\caption{\label{fig:Results2}
	(Upper panel) the $\chi^{2}-$like test for the quasienergy spacings $s$ and (lower panel) the spread of the distribution of the decay widths $\Gamma$.
	The size of the system is $N=8$, $L=7$.
	The dashed lines are a guide to the eye and suggest that the transition to the chaotic regime can be appreciated by looking at both, the real and the imaginary parts of the FB eigenvalues.
	}
\end{figure}
The total width $\Gamma_{\rm F}=\Gamma_{1}+\Gamma_{2}$ is now added to the single-band Hamiltonian as a complex shift.
Given the translational symmetry of this Hamiltonian, we added the shift to the diagonal in the representation of the cyclic basis $|\sigma\kappa\rangle$ in Eq.~\eqref{eq:FloquetBloch2}, as $-i\Gamma_{\rm F}(\sigma)/2$.
The eigenvalues of the FB operator are no longer unitary and the quasienergies $E_{j}$ has a complex part $-i\Gamma_{j}/2$.
We analyzed the decay rates $\Gamma_{j}$ \emph{along} with the quasienergy spacings statistics to study how the dynamics within the first band influences the coupling to the second band.
We first reported the distribution $\mathsf{P}(\Gamma)$ in \cite{TMW2007}.
Here we refine our analysis and we first focus on $\mathsf{P}(\Lambda)$, with $\Lambda=\log_{10}\Gamma$, shown in Fig.~\ref{fig:Results1} for some paradigmatic cases.
The widths are much smaller than unity, consistently with a perturbative approach of the system, yet the lattice potential is only $V=1.5E_{R}$ to increase the spread of the Stark state in the second band.
Moreover, the decay channels are activated by an increase of the interaction energy, which can be experimentally achieved by acting on the transversal confining potential \cite{Group12} of the quasi-one dimensional lattice, or by a Feshbach resonance \cite{BDZ2007}. 
In this Section, $W\simeq 0.02$ used in \cite{BK2003} is multiplied by a factor of order $10$, a value that is still well within the experimental possibilities \cite{BDZ2007,KMSGE2005}.
\begin{figure} 
  \centering 
  \includegraphics[width=\columnwidth]{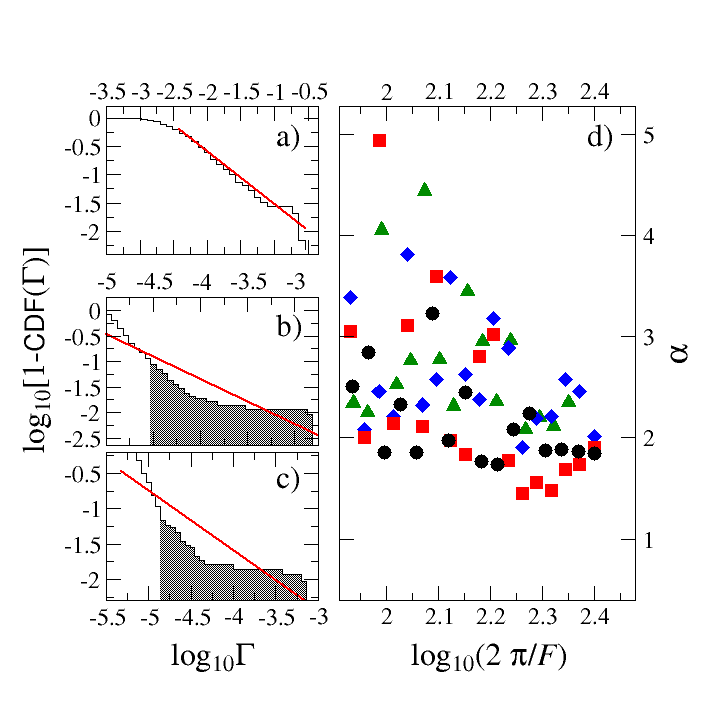}
	\caption{\label{fig:Results3}
	(Color online)
	(a-c) l.h.s of Eq.~\eqref{eq:Results1} in the regular (a) and in the chaotic regime (b,c), for a system with $N=8$, $L=7$.
	The dashed area in panels (b,c) shows the part of the histogram where the total amount of points is about $40$, less then $10$\% of the full sample .
	The red solid line is the linear fit to the profile in the scaling region.
	(d) The exponent of the power law  $\mathsf{P}(\Gamma)\sim\Gamma^{-\alpha}$ as obtained from the linear fits shown in (a-c), for $N=8$, $L=7$ (black circles),  $N=8$, $L=5$ (red squares),  $N=7$, $L=8$ (blue diamonds), $N=7$, $L=5$ (green pyramids). 
	To show together different data sets we horizontally translated the first by $+0.1$ and the last by $-0.05$.
	}
\end{figure}

In Fig.~\ref{fig:Results1} we compare the distribution of the decay widths for two values of $F$ that belong to the regular (a,c) and the chaotic region (b,d).
The difference in the average decay width $\langle \Lambda \rangle$ is due to the improved energy matching provided by a stronger external field $F$, that supplies the necessary energy to promote particles to the second band.
For the parameters of Fig.~\ref{fig:Results1}, the single-particle Landau-Zener formula \cite{GKK2002} gives 
$\Gamma_{\mathrm{LZ}} = F/(2\pi) {\rm exp} [ -\pi^2 (\varepsilon_{2}-\varepsilon_{1})^{2} /(8 F) ]
\sim 10^{-20}$, $10^{-75}$ for (a,b).
Then we see that there are regimes where the many-body interactions affect substantially the single-particle tunneling rates and {\it cannot} be neglected.
Moreover, even mean-field treatments of the Landau-Zener tunneling \cite{SZLWCMA2007,Group10} cannot account for several decay channels.

The \emph{bulks} of the distributions, both in the regular and in the chaotic regime, are appropriately fit by a lognormal profile $\mathsf{P}(\Lambda)\propto {\rm exp}\{-[\ln (\Lambda-\Lambda_{\rm min})]^{2}/2\Delta\Lambda^{2}\}/(\Lambda-\Lambda_{\rm min})$.
We notice that the spread $\Delta\Lambda$ is reduced in the chaotic case, where the Fock states are strongly mixed by the dynamics and a fast decaying Fock state (in the $|\sigma\kappa\rangle$ representation) could be a privileged decay channel for \emph{many} eigenstates of the system. 
Many eigenstates then shared similar decay widths and their statistical distribution would be thinner.
Following this reasoning, we interpret the thinner distributions found in the panels (b,d) of Fig.~\ref{fig:Results1} as a signature of the strongly correlated dynamics.

This picture is supported by Fig.~\ref{fig:Results2}, where the dependence on $F$ of the spread $\Delta\Lambda$ (lower panel) is confronted with the regular-chaotic transition evidenced by the $\chi^{2}-$like test of Eq.~\eqref{eq:oneBand6} (upper panel).
The shrinking of the decay widths distribution goes along with the transition, though the precise determination of a transition point is precluded by a substantial amount of noise.
The finite size of the samples that can be managed numerically accounts for the noise, as we verified that
the profile becomes sharper while increasing the size of the system.
Moreover, since the average decay width decreases with smaller $F$, we need a more precise and hence a more time-expensive numerical computation to determine $\Delta\Lambda$ as we enter the chaotic regime.

Finally, we found in \cite{TMW2007} that the tails of the distributions in the chaotic case follow the expected power-law for the diffusive regime of an open quantum chaotic system. 
In our case, the opening of the ground band subsystem is defined by the interband coupling, which in a sense attaches ``leads'' to {\em all} lattice sites {\em within} the sample. 
Indeed, the inset of Fig.~\ref{fig:Results1} (d) shows that the lognormal is not a good fit to the tails in the chaotic regime: the distributions transform to a power-law $\mathsf{P}(\Gamma)\sim\Gamma^{-\alpha}$ in close analogy to the transition from Anderson-localized to diffusive dynamics in open disordered systems \cite{Kottos2005,Group21}. 
In particular we found $\alpha \simeq 2$, in accordance with the general results of \cite{FS1996}.
In the Fig.~\ref{fig:Results3}, panel (d) we show that this value is indeed peculiar to the chaotic regime, and strong fluctuations with $F$ mean that the exponent $\alpha$ is not defined within the transition region.
Here we evaluate the integrated profile
\begin{equation}\label{eq:Results1}
1-\mathsf{CDF}(\Gamma) \sim 1-\int^{\Gamma}\mathsf{P}(\Gamma'){\rm d}\Gamma' \sim \Gamma^{1-\alpha},
\end{equation}
and fit the latter to reduce the statistical fluctuations due to the finiteness of our samples.
The finite-size effects are more relevant in the chaotic case (panels b,c), because the reduced spread of the distribution makes the fit sensible to the few points that fall in the further part of the tail (shaded in the pictures).
The largest analyzed sample has only $\simeq 400$ energy levels, yet the uncertainty on $\alpha\simeq 2$ on the chaotic side of the transition is less than $10$\%.

\section{\label{sec:Conclusion} Conclusions}

We studied the problem of a many-body atomic system in a tilted optical lattice, using a statistical analysis of the complete quantum spectrum.
We extended previous work \cite{KB2003,BK2003,TMW2007} and verified thoroughly the transition from a regular to a quantum chaotic spectrum, found in \cite{KB2003} as the Stark field is varied.
The transition takes place when the tunneling amplitude in the lattice becomes comparable to the interaction energy of the atoms. 
In this regime it is not possible to find a set of quantum numbers that decompose the spectrum into subspaces not mixed by a change of the Stark field. 
Because of the strong mixing of the energy eigenstates, many avoided crossings are found in the spectrum and the statistical analysis of the energy spacings show a characteristic depletion of small spacings - a signature of quantum chaos.

We derived a reasonable two-bands model, on the basis of which we opened the ground band subsystem by a perturbative coupling of the Fock states to excited Stark states.
We obtained the set of decay rates from the complex-valued quasienergies of the FB operator.
We analyzed the real part of the quasienergies and verified that the transition from the regular to the chaotic regime is still visible and not much modified with respect to the one-band system.
Moreover we analyzed the statistical distribution of the imaginary part of the quasienergies, i.e. the decay rates of the states in the ground band.
We found that the distributions of the decay rates become thinner as the regular-chaotic transition is crossed.
We reckon that thinner distributions of the decay rates are a signature of the complex dynamics, where a few strongly decaying states act as leading decay channels.
The statistical characterization of the tunneling rates could be used to compute the expected atomic current from the ground band to the excited band of the lattice, thus providing an experimental probe for the regular-chaotic transition.
Time-dependent observables could possibly be computed also with advanced mean-field techniques as reported in Ref.~\cite{Cederbaum2}, as long as no full spectral information of the many-body system is wanted.

Of course, a deeper investigation is desirable to understand on quantitative grounds the full interplay of the dynamics within the first band and the decay towards higher bands, and a more detailed analysis of the interband coupling will be worthwhile in a full-blown model in which at least two bands are completely included. 
Our results are a first step in this direction of studies of regimes in which both ``horizontal'' and ``vertical'' quantum transport are simultaneously present and influence each other in a complex manner.

\begin{acknowledgments}
We thank Alexey Ponomarev, Gregor Veble, and Andrey Kolovsky for lively discussions on the Wannier-Stark system and Quantum Chaos and Yan Fyodorov for bringing Ref.~\cite{FS1996} to our attention. 
We are grateful to the Centro di Calcolo, Dipartimento di Fisica, Universit\`a di Pisa, for providing CPU and kind assistance, and for support by MIUR-PRIN and EU-OLAQUI. 
S.~W. furthermore acknowledges support within the framework of the Excellence Initiative by the German Research Foundation (DFG) through the Heidelberg Graduate School of Fundamental Physics (grant number GSC 129/1).
\end{acknowledgments}

\appendix{}
\section{\label{sec:Wannier} Computation of the coefficients of the many-body Hamiltonian}

The coefficients of the Hamiltonian Eq.~\eqref{eq:oneBand3} are given by integrals of the WF $\chi_{\ell,\alpha }$, once the expansion of Eq.~\eqref{eq:oneBand2} is substituted into Eq.~\eqref{eq:oneBand4}. 
The WF are defined as:
\begin{equation}\label{eq:Wannier1}
\psi_{k,\alpha }(x)=\sqrt{\frac{1}{2\pi} } \sum_{\ell } e^{i\pi \ell k} \chi_{\ell,\alpha }(x).
\end{equation}
From the definition it follows that $\chi_{\ell,\alpha}(x)=\chi_{\alpha}(x-\ell a)$, so that all the Bloch waves for a band can be computed starting from a \emph{single} Wannier function $\chi_{\alpha}$.
We computed the WF following the method introduced in \cite{Slater1952}.
A Bloch wave is factorized as $\psi_{k,\alpha }(x)=e^{ikx}u_{k,\alpha}$, where the periodic function $u_{k,\alpha }(x)=u_{k,\alpha}(x+a)$ is expanded in a truncated basis of momenta using multiples of $2\pi j/a$ only:
\begin{equation}\label{eq:Wannier2}
u_{k,\alpha }(x)=\sum_{j=-Q}^{Q-1}u_{j}(\alpha, k) \frac{1}{\sqrt{a}} e^{i2 jx}.
\end{equation}
Using a gauge transform ($p\rightarrow -i\partial_{x} + k$), an effective Hamiltonian for the periodic $u_{k,\alpha}$ is derived from Eq.~\eqref{eq:oneBand1}.
We solved the effective Schr\"odinger equation as a linear system for $u_{j}(k,\alpha)$ with dispersion law $E_{k,\alpha}$ as eigenvalues.
We obtained the on-site energies and the hopping amplitudes from the Fourier transform of $E_{k,\alpha}$:
\begin{equation}\label{eq:Wannier3}
\varepsilon_{\alpha}=\frac{1}{2}\int_{-1}^{+1}E_{k,\alpha}{\rm d}k, \quad
J_{\alpha}=-\int_{-1}^{+1}E_{k,\alpha}e^{i\pi k}{\rm d}k.
\end{equation}
The WF is finally computed from the inversion of Eq.~\eqref{eq:Wannier1}.
The relative phase of the eigenvectors $u_{j}(k,\alpha)$, for different quasimomenta, is \emph{not} unique.
For the simple case of a sinusoidal lattice potential, the correct choice of the phases can be inferred from \cite{Slater1952}:
\begin{equation}
u_{j}(k,1)\,\rightarrow\, |u_{j}(k,1)|,\;
u_{j}(k,2)\,\rightarrow\, |u_{j}(k,2)| {\rm sign} (2j+k).
\end{equation}
This phase choice guarantees the correct inversion symmetry $\chi_{\alpha}(-x) = (-1)^{\alpha-1}\chi_{\alpha}(x)$ of the WF of the first and of the second band, respectively.
The interaction coefficients for Eq.~\eqref{eq:oneBand3} read:
\begin{equation}\label{eq:Wannier4}
W_{\alpha}=W \int_{-\infty}^{\infty}\chi_{\alpha}^{4}\;{\rm d}x,\quad
W_{\times} = W \int_{-\infty}^{\infty}\chi_{1}^{2}\chi_{2}^{2}\;{\rm d}x.
\end{equation}
The WF are an orthogonal set of functions, so that different bands are decoupled in the one-body dynamics.
The tilting potential $Fx$ has nonvanishing ``dipole'' matrix elements that couples adjacent bands:
\begin{equation}\label{eq:Wannier5}
d_{F} = \int_{-\infty}^{\infty}\chi_{1}(x)x\chi_{2}(x)\;{\rm d}x.
\end{equation}

\section{\label{sec:FloquetBloch} Computation of the Floquet-Bloch operator}

The Fock states $|\vec{n}\rangle$ are defined by the sequence of occupation numbers $\{n_{\ell,\alpha}\}_{\ell=1}^{L}$ for the $L$ sites of the lattice, in each band $\alpha$.
The cyclic boundary conditions, $n_{0,\alpha }  = n_{L,\alpha }$ allow us to define a shift operator $\hat{S}$ \cite{KB2003} that translates the occupation numbers of a Fock state $|n \rangle$:
\begin{equation}\label{eq:FloquetBloch1}
\hat{S}^{m} |n \rangle =| \dots,n_{1-m,\alpha },\dots,n_{L-m,\alpha },\dots\rangle.
\end{equation}
The operator $\hat{S}$ naturally decomposes the Fock space into equivalence classes of vectors generated by repeated application of $\hat{S}$ onto a ``seed'' vector $|\sigma \rangle$ with $M(\sigma)\le L$ such that $\hat{S}^{M(\sigma)}|n \rangle = |n \rangle$  for all the vectors $|n\rangle$ in the class.
A new basis $|\sigma \kappa \rangle$ can be introduced for which the many-body quasimomentum $\kappa = j / M(\sigma)$ ($0\le j <M(\sigma)$), supplies a convenient label:
\begin{equation}\label{eq:FloquetBloch2}
| \sigma \kappa \rangle =\frac{1}{\sqrt{M(\sigma)}} \sum_{\ell=1}^{M(\sigma)} e^{i2 \pi \kappa \ell } \hat{S}^{\ell} |\sigma\rangle.
\end{equation}
The cyclic basis decomposes the Hamiltonian Eq.~\eqref{eq:oneBand3} into a block form that transfers to the FB operator, whence $\hat {\cal U}_{\rm FB}=\oplus_{j=1}^{L}\; \hat{\cal U}_{\rm FB}(\kappa=j/L)$ with the obvious advantage that we can diagonalize separately each block of dimension $D\lesssim D_{\rm H}/L$.
This decomposition not only leads to a substantial numerical simplification but is also of dynamical relevance \cite{KB2003}.
Moreover, we exploited that the Hamiltonian matrix is \emph{sparse} and we verified that the fraction of the nonzero entries is $4.0 \times D^{-1.1}$ in the limit of large Hilbert spaces $D\gg 1$.

The column $c$ of the FB operator coincides with the column of the coefficients of the basis state of index $c$, evolved in time up to one Bloch period.
We used a fourth-order Runge-Kutta time integrator with adaptive stepsize, tuned in precision by the upper bound $\varepsilon$ of the estimated one-step error \cite{NumericalRecipes}.
The value of $\varepsilon$ was chosen to suppress, up to $T_{\rm B}$, the well-known exponential instability of the Runge-Kutta method applied to the Schr\"odinger equation.
The quantity $Q(E,E^{\prime})\equiv \left (\sum_{i}|E_{i} - E'_{i}|^{2}/D^{2} \right )^{1/2}$ was used to compare different spectra $\{E \}$, $\{ E' \}$ and we verified the consistency of our computations, finding a power-law self-convergence  $Q(E^{(r)},E^{(r-1)})\propto \varepsilon_{r}^{1.2}$ for a sequence of tests with increasing required precision, $\{\varepsilon_{r}\}_{r}\rightarrow 0$.
The achieved precision scales with cpu time $t$ as $Q\propto t^{-6.1}$.
A precision up to $10^{-11}$ was necessary to reliably compute the small complex part in the eigenvalues of the FB operator, analyzed in Section~\ref{sec:Results}.


\begin{thebibliography}{35}
\expandafter\ifx\csname natexlab\endcsname\relax\def\natexlab#1{#1}\fi
\expandafter\ifx\csname bibnamefont\endcsname\relax
  \def\bibnamefont#1{#1}\fi
\expandafter\ifx\csname bibfnamefont\endcsname\relax
  \def\bibfnamefont#1{#1}\fi
\expandafter\ifx\csname citenamefont\endcsname\relax
  \def\citenamefont#1{#1}\fi
\expandafter\ifx\csname url\endcsname\relax
  \def\url#1{\texttt{#1}}\fi
\expandafter\ifx\csname urlprefix\endcsname\relax\def\urlprefix{URL }\fi
\providecommand{\bibinfo}[2]{#2}
\providecommand{\eprint}[2][]{\url{#2}}

\bibitem[{Gro({\natexlab{a}})}]{Group1}
\bibinfo{author}{\bibfnamefont{J.}~\bibnamefont{Hecker-Denschlag}},
  \bibinfo{author}{\bibfnamefont{J.~E.} \bibnamefont{Simsarian}},
  \bibinfo{author}{\bibfnamefont{H.}~\bibnamefont{{H\"affner}}},
  \bibinfo{author}{\bibfnamefont{C.}~\bibnamefont{McKenzie}},
  \bibinfo{author}{\bibfnamefont{A.}~\bibnamefont{Browaeys}},
  \bibinfo{author}{\bibfnamefont{D.}~\bibnamefont{Cho}},
  \bibinfo{author}{\bibfnamefont{K.}~\bibnamefont{Helmerson}},
  \bibinfo{author}{\bibfnamefont{S.~L.} \bibnamefont{Rolston}},
  \bibnamefont{and} \bibinfo{author}{\bibfnamefont{W.~D.}
  \bibnamefont{Phillips}}, \bibinfo{journal}{J.\ Phys.\ B}
  \textbf{\bibinfo{volume}{35}}, \bibinfo{pages}{3095} (\bibinfo{year}{2002});
\bibinfo{author}{\bibfnamefont{I.}~\bibnamefont{Bloch}}, 
  \bibinfo{journal}{J.\ Phys.\ B} \textbf{\bibinfo{volume}{38}},
  \bibinfo{pages}{S625} (\bibinfo{year}{2005});
\bibinfo{author}{\bibfnamefont{O.}~\bibnamefont{Morsch}} \bibnamefont{and}
  \bibinfo{author}{\bibfnamefont{M.}~\bibnamefont{Oberthaler}},
  \bibinfo{journal}{Rev.\ Mod.\ Phys.} \textbf{\bibinfo{volume}{78}},
  \bibinfo{pages}{179} (\bibinfo{year}{2006}).

\bibitem[{\citenamefont{Scully and Rempe}(2006)}]{SR2006}
\bibinfo{author}{\bibfnamefont{M.}~\bibnamefont{Scully}} \bibnamefont{and}
  \bibinfo{author}{\bibfnamefont{G.}~\bibnamefont{Rempe}},
  \emph{\bibinfo{title}{Advances in Atomic, Molecular, and Optical Physics,
  53}} (\bibinfo{publisher}{Academic Press}, \bibinfo{address}{Amsterdam},
  \bibinfo{year}{2006}).

\bibitem[{Gro({\natexlab{b}})}]{Group2}
\bibinfo{author}{\bibfnamefont{E.~E.} \bibnamefont{Mendez}},
  \bibinfo{author}{\bibfnamefont{F.}~\bibnamefont{Agullo-Rueda}},
  \bibnamefont{and} \bibinfo{author}{\bibfnamefont{J.~M.} \bibnamefont{Hong}},
  \bibinfo{journal}{Phys.\ Rev.\ Lett.} \textbf{\bibinfo{volume}{60}},
  \bibinfo{pages}{2426} (\bibinfo{year}{1988});
\bibinfo{author}{\bibfnamefont{J.}~\bibnamefont{Feldmann}},
  \bibinfo{author}{\bibfnamefont{K.}~\bibnamefont{Leo}}, 
  \bibinfo{author}{\bibfnamefont{J.}~\bibnamefont{Shah}},
  \bibinfo{author}{\bibfnamefont{D.~A.~B.}~\bibnamefont{Miller}},
  \bibinfo{author}{\bibfnamefont{J.~E.}~\bibnamefont{Cunningham}},
  \bibinfo{author}{\bibfnamefont{T.}~\bibnamefont{Meier}},
  \bibinfo{author}{\bibfnamefont{G.}~\bibnamefont{{von~Plessen}}},
  \bibinfo{author}{\bibfnamefont{A.}~\bibnamefont{Schulze}},
  \bibinfo{author}{\bibfnamefont{P.}~\bibnamefont{Thomas}} \bibnamefont{and}
  \bibinfo{author}{\bibfnamefont{S.}~\bibnamefont{{Schmitt-Rink}}},
  \bibinfo{journal}{Phys.\ Rev.\ B} \textbf{\bibinfo{volume}{46}},
  \bibinfo{pages}{7252} (\bibinfo{year}{1992});
\bibinfo{author}{\bibfnamefont{K.}~\bibnamefont{Leo}},
  \emph{\bibinfo{title}{High-Field Transport in Semiconductor Superlattices}}
  (\bibinfo{publisher}{Springer}, \bibinfo{address}{Berlin},
  \bibinfo{year}{2003}).

\bibitem[{Gro({\natexlab{c}})}]{Group3}
\bibinfo{author}{\bibfnamefont{T.}~\bibnamefont{Pertsch}},
  \bibinfo{author}{\bibfnamefont{P.}~\bibnamefont{Dannberg}},
  \bibinfo{author}{\bibfnamefont{W.}~\bibnamefont{Elflein}},
  \bibinfo{author}{\bibfnamefont{A.}~\bibnamefont{{Br\"auer}}},
  \bibnamefont{and} \bibinfo{author}{\bibfnamefont{F.}~\bibnamefont{Lederer}},
  \bibinfo{journal}{Phys.\ Rev.\ Lett.} \textbf{\bibinfo{volume}{83}},
  \bibinfo{pages}{4752} (\bibinfo{year}{1999});
\bibinfo{author}{\bibfnamefont{R.}~\bibnamefont{Morandotti}},
  \bibinfo{author}{\bibfnamefont{U.}~\bibnamefont{Peschel}},
  \bibinfo{author}{\bibfnamefont{J.~S.}~\bibnamefont{Aitchison}},
  \bibinfo{author}{\bibfnamefont{H.~S.}~\bibnamefont{Eisenberg}} \bibnamefont{and}
  \bibinfo{author}{\bibfnamefont{Y.}~\bibnamefont{Silberberg}},
  \bibinfo{journal}{Phys.\ Rev.\ Lett.} \textbf{\bibinfo{volume}{83}},
  \bibinfo{pages}{4756} (\bibinfo{year}{1999});
\bibinfo{author}{\bibfnamefont{M.}~\bibnamefont{Ghulinyan}},
  \bibinfo{author}{\bibfnamefont{C.~J.} \bibnamefont{Oton}},
  \bibinfo{author}{\bibfnamefont{Z.}~\bibnamefont{Gaburro}},
  \bibinfo{author}{\bibfnamefont{L.}~\bibnamefont{Pavesi}},
  \bibinfo{author}{\bibfnamefont{C.}~\bibnamefont{Toninelli}},
  \bibnamefont{and} \bibinfo{author}{\bibfnamefont{D.~S.}~\bibnamefont{Wiersma}},
  \bibinfo{journal}{Phys.\ Rev.\ Lett.} \textbf{\bibinfo{volume}{94}},
  \bibinfo{pages}{127401} (\bibinfo{year}{2005}).

\bibitem[{Gro({\natexlab{d}})}]{Group4}
\bibinfo{author}{\bibfnamefont{M.}~\bibnamefont{BenDahan}},
  \bibinfo{author}{\bibfnamefont{E.}~\bibnamefont{Peik}},
  \bibinfo{author}{\bibfnamefont{J.}~\bibnamefont{Reichel}},
  \bibinfo{author}{\bibfnamefont{Y.}~\bibnamefont{Castin}}, \bibnamefont{and}
  \bibinfo{author}{\bibfnamefont{C.}~\bibnamefont{Salomon}},
  \bibinfo{journal}{Phys.\ Rev.\ Lett.} \textbf{\bibinfo{volume}{76}},
  \bibinfo{pages}{4508} (\bibinfo{year}{1996});
\bibinfo{author}{\bibfnamefont{S.~R.} \bibnamefont{Wilkinson}},
  \bibinfo{author}{\bibfnamefont{C.~F.} \bibnamefont{Bharucha}},
  \bibinfo{author}{\bibfnamefont{K.~W.} \bibnamefont{Madison}},
  \bibinfo{author}{\bibfnamefont{Q.}~\bibnamefont{Niu}}, \bibnamefont{and}
  \bibinfo{author}{\bibfnamefont{M.~G.} \bibnamefont{Raizen}},
  \bibinfo{journal}{Phys.\ Rev.\ Lett.} \textbf{\bibinfo{volume}{76}},
  \bibinfo{pages}{4512} (\bibinfo{year}{1996});
\bibinfo{author}{\bibfnamefont{B.~P.} \bibnamefont{Anderson}} \bibnamefont{and}
  \bibinfo{author}{\bibfnamefont{M.~A.} \bibnamefont{Kasevich}},
  \bibinfo{journal}{Science} \textbf{\bibinfo{volume}{282}},
  \bibinfo{pages}{1686} (\bibinfo{year}{1998}).

\bibitem[{Gro({\natexlab{e}})}]{Group5}
\bibinfo{author}{\bibfnamefont{O.}~\bibnamefont{Morsch}},
  \bibinfo{author}{\bibfnamefont{M.}~\bibnamefont{Cristiani}},
  \bibinfo{author}{\bibfnamefont{J.~H.} \bibnamefont{{M\"uller}}},
  \bibinfo{author}{\bibfnamefont{D.}~\bibnamefont{Ciampini}}, \bibnamefont{and}
  \bibinfo{author}{\bibfnamefont{E.}~\bibnamefont{Arimondo}},
  \bibinfo{journal}{Phys.\ Rev.\ A} \textbf{\bibinfo{volume}{66}},
  \bibinfo{pages}{021601(R)} (\bibinfo{year}{2002});
\bibinfo{author}{\bibfnamefont{O.}~\bibnamefont{Morsch}},
  \bibinfo{author}{\bibfnamefont{J.~H.} \bibnamefont{{M\"uller}}},
  \bibinfo{author}{\bibfnamefont{M.}~\bibnamefont{Cristiani}},
  \bibinfo{author}{\bibfnamefont{D.}~\bibnamefont{Ciampini}}, \bibnamefont{and}
  \bibinfo{author}{\bibfnamefont{E.}~\bibnamefont{Arimondo}},
  \bibinfo{journal}{Phys.\ Rev.\ Lett.} \textbf{\bibinfo{volume}{87}},
  \bibinfo{pages}{140402} (\bibinfo{year}{2001}).

\bibitem[{\citenamefont{Roati et~al.}(2004)\citenamefont{Roati, {de~Mirandes},
  Ferlaino, Ott, Modugno, and Inguscio}}]{RMFOMI2004}
\bibinfo{author}{\bibfnamefont{G.}~\bibnamefont{Roati}},
  \bibinfo{author}{\bibfnamefont{E.}~\bibnamefont{{de~Mirandes}}},
  \bibinfo{author}{\bibfnamefont{F.}~\bibnamefont{Ferlaino}},
  \bibinfo{author}{\bibfnamefont{H.}~\bibnamefont{Ott}},
  \bibinfo{author}{\bibfnamefont{G.}~\bibnamefont{Modugno}}, \bibnamefont{and}
  \bibinfo{author}{\bibfnamefont{M.}~\bibnamefont{Inguscio}},
  \bibinfo{journal}{Phys.\ Rev.\ Lett.} \textbf{\bibinfo{volume}{92}},
  \bibinfo{pages}{230402} (\bibinfo{year}{2004}).

\bibitem{SZLWCMA2007}
\bibinfo{author}{\bibfnamefont{C.}~\bibnamefont{Sias}},
  \bibinfo{author}{\bibfnamefont{A.}~\bibnamefont{Zenesini}},
  \bibinfo{author}{\bibfnamefont{H.}~\bibnamefont{Lignier}},
  \bibinfo{author}{\bibfnamefont{S.}~\bibnamefont{Wimberger}},
  \bibinfo{author}{\bibfnamefont{D.}~\bibnamefont{Ciampini}},
  \bibinfo{author}{\bibfnamefont{O.}~\bibnamefont{Morsch}} \bibnamefont{and}
  \bibinfo{author}{\bibfnamefont{E.}~\bibnamefont{Arimondo}},
  \bibinfo{journal}{Phys.\ Rev.\ Lett.} \textbf{\bibinfo{volume}{98}},
  \bibinfo{pages}{120403} (\bibinfo{year}{2007}).

\bibitem[{\citenamefont{Buchleitner and Kolovsky}(2003)}]{BK2003}
\bibinfo{author}{\bibfnamefont{A.}~\bibnamefont{Buchleitner}} \bibnamefont{and}
  \bibinfo{author}{\bibfnamefont{A.~R.} \bibnamefont{Kolovsky}},
  \bibinfo{journal}{Phys.\ Rev.\ Lett.} \textbf{\bibinfo{volume}{91}},
  \bibinfo{pages}{253002} (\bibinfo{year}{2003}).

\bibitem[{\citenamefont{Kolovsky and Buchleitner}(2003)}]{KB2003}
\bibinfo{author}{\bibfnamefont{A.~R.} \bibnamefont{Kolovsky}} \bibnamefont{and}
  \bibinfo{author}{\bibfnamefont{A.}~\bibnamefont{Buchleitner}},
  \bibinfo{journal}{Phys.\ Rev.\ E} \textbf{\bibinfo{volume}{68}},
  \bibinfo{pages}{056213} (\bibinfo{year}{2003}).

\bibitem[{\citenamefont{Kolovsky}(2003)}]{Kolovsky2003}
\bibinfo{author}{\bibfnamefont{A.~R.} \bibnamefont{Kolovsky}},
  \bibinfo{journal}{Phys.\ Rev.\ Lett.} \textbf{\bibinfo{volume}{90}},
  \bibinfo{pages}{213002} (\bibinfo{year}{2003}).

\bibitem[{\citenamefont{Ponomarev et~al.}(2006)\citenamefont{Ponomarev,
  {Madro\~nero}, Kolovsky, and Buchleitner}}]{PMKB2006}
\bibinfo{author}{\bibfnamefont{A.~V.} \bibnamefont{Ponomarev}},
  \bibinfo{author}{\bibfnamefont{J.}~\bibnamefont{{Madro\~nero}}},
  \bibinfo{author}{\bibfnamefont{A.~R.} \bibnamefont{Kolovsky}},
  \bibnamefont{and}
  \bibinfo{author}{\bibfnamefont{A.}~\bibnamefont{Buchleitner}},
  \bibinfo{journal}{Phys.\ Rev.\ Lett.} \textbf{\bibinfo{volume}{96}},
  \bibinfo{pages}{050404} (\bibinfo{year}{2006}).
  
\bibitem[{Gro({\natexlab{f}})}]{Group6}
\bibinfo{author}{\bibfnamefont{M.}~\bibnamefont{Greiner}},
  \bibinfo{author}{\bibfnamefont{O.}~\bibnamefont{Mandel}},
  \bibinfo{author}{\bibfnamefont{T.}~\bibnamefont{Esslinger}},
  \bibinfo{author}{\bibfnamefont{T.~W.} \bibnamefont{{H\"ansch}}},
  \bibnamefont{and} \bibinfo{author}{\bibfnamefont{I.}~\bibnamefont{Bloch}},
  \bibinfo{journal}{Nature} \textbf{\bibinfo{volume}{415}}, \bibinfo{pages}{39}
  (\bibinfo{year}{2002});
\bibinfo{author}{\bibfnamefont{O.}~\bibnamefont{Mandel}},
  \bibinfo{author}{\bibfnamefont{M.}~\bibnamefont{Greiner}},
  \bibinfo{author}{\bibfnamefont{A.}~\bibnamefont{Widera}},
  \bibinfo{author}{\bibfnamefont{T.}~\bibnamefont{Rom}},
  \bibinfo{author}{\bibfnamefont{T.~W.} \bibnamefont{{H\"ansch}}},
  \bibnamefont{and} \bibinfo{author}{\bibfnamefont{I.}~\bibnamefont{Bloch}},
  \bibinfo{journal}{Nature} \textbf{\bibinfo{volume}{425}},
  \bibinfo{pages}{937} (\bibinfo{year}{2003});
\bibinfo{author}{\bibfnamefont{F.}~\bibnamefont{Gerbier}},
  \bibinfo{author}{\bibfnamefont{S.}~\bibnamefont{{F\"olling}}},
  \bibinfo{author}{\bibfnamefont{A.}~\bibnamefont{Widera}},
  \bibinfo{author}{\bibfnamefont{O.}~\bibnamefont{Mandel}}, \bibnamefont{and}
  \bibinfo{author}{\bibfnamefont{I.}~\bibnamefont{Bloch}},
  \bibinfo{journal}{Phys.\ Rev.\ Lett.} \textbf{\bibinfo{volume}{96}},
  \bibinfo{pages}{090401} (\bibinfo{year}{2006}).

\bibitem{BDZ2007}
\bibinfo{author}{\bibfnamefont{I.}~\bibnamefont{Bloch}},
  \bibinfo{author}{\bibfnamefont{J.}~\bibnamefont{Dalibard}},
  \bibnamefont{and} \bibinfo{author}{\bibfnamefont{W.} \bibnamefont{Zwerger}}
  (\bibinfo{year}{2007}), \eprint{cond-mat.other/0704.3011v1}.

\bibitem[{Gro({\natexlab{g}})}]{Group9}
\bibinfo{author}{\bibfnamefont{P.}~\bibnamefont{Buonsante}},
  \bibinfo{author}{\bibfnamefont{R.}~\bibnamefont{Franzosi}}, \bibnamefont{and}
  \bibinfo{author}{\bibfnamefont{V.}~\bibnamefont{Penna}},
  \bibinfo{journal}{Phys.\ Rev.\ Lett.} \textbf{\bibinfo{volume}{90}},
  \bibinfo{pages}{050404} (\bibinfo{year}{2003});
  \bibinfo{journal}{J.\ Phys.\ B} \textbf{\bibinfo{volume}{37}},
  \bibinfo{pages}{S229} (\bibinfo{year}{2004});
\bibinfo{author}{\bibfnamefont{M.}~\bibnamefont{Hiller}},
  \bibinfo{author}{\bibfnamefont{T.}~\bibnamefont{Kottos}}, \bibnamefont{and}
  \bibinfo{author}{\bibfnamefont{T.}~\bibnamefont{Geisel}},
  \bibinfo{journal}{Phys.\ Rev.\ A} \textbf{\bibinfo{volume}{73}},
  \bibinfo{pages}{061604(R)} (\bibinfo{year}{2006}).

\bibitem[{\citenamefont{Tomadin et~al.}(2007)\citenamefont{Tomadin, Mannella,
  and Wimberger}}]{TMW2007}
\bibinfo{author}{\bibfnamefont{A.}~\bibnamefont{Tomadin}},
  \bibinfo{author}{\bibfnamefont{R.}~\bibnamefont{Mannella}}, \bibnamefont{and}
  \bibinfo{author}{\bibfnamefont{S.}~\bibnamefont{Wimberger}},
  \bibinfo{journal}{Phys.\ Rev.\ Lett.} \textbf{\bibinfo{volume}{98}},
  \bibinfo{pages}{130402} (\bibinfo{year}{2007}).

\bibitem[{\citenamefont{Fisher et~al.}(1989)\citenamefont{Fisher, Weichman,
  Grinstein, and Fisher}}]{FWGF1989}
\bibinfo{author}{\bibfnamefont{M.~P.~A.} \bibnamefont{Fisher}},
  \bibinfo{author}{\bibfnamefont{P.~B.} \bibnamefont{Weichman}},
  \bibinfo{author}{\bibfnamefont{G.}~\bibnamefont{Grinstein}},
  \bibnamefont{and} \bibinfo{author}{\bibfnamefont{D.~S.}~\bibnamefont{Fisher}},
  \bibinfo{journal}{Phys.\ Rev.\ B} \textbf{\bibinfo{volume}{40}},
  \bibinfo{pages}{546} (\bibinfo{year}{1989}).

\bibitem[{\citenamefont{Sachdev et~al.}(2002)\citenamefont{Sachdev, Sengupta,
  and Girvin}}]{SSG2002}
\bibinfo{author}{\bibfnamefont{S.}~\bibnamefont{Sachdev}},
  \bibinfo{author}{\bibfnamefont{K.}~\bibnamefont{Sengupta}}, \bibnamefont{and}
  \bibinfo{author}{\bibfnamefont{S.~M.} \bibnamefont{Girvin}},
  \bibinfo{journal}{Phys.\ Rev.\ B} \textbf{\bibinfo{volume}{66}},
  \bibinfo{pages}{075128} (\bibinfo{year}{2002}).

\bibitem[{\citenamefont{Scarola and {Das~Sarma}}(2005)}]{SS2005}
\bibinfo{author}{\bibfnamefont{V.~W.} \bibnamefont{Scarola}} \bibnamefont{and}
  \bibinfo{author}{\bibfnamefont{S.}~\bibnamefont{{Das~Sarma}}},
  \bibinfo{journal}{Phys.\ Rev.\ Lett.} \textbf{\bibinfo{volume}{95}},
  \bibinfo{pages}{033003} (\bibinfo{year}{2005}).

\bibitem[{Gro({\natexlab{h}})}]{Group7}
\bibinfo{author}{\bibfnamefont{W.}~\bibnamefont{Zwerger}},
  \bibinfo{journal}{J.\ Opt.\ B: Quant.\ S.\ Opt.}
  \textbf{\bibinfo{volume}{5}}, \bibinfo{pages}{S9} (\bibinfo{year}{2003});
\bibinfo{author}{\bibfnamefont{R.}~\bibnamefont{Roth}} \bibnamefont{and}
  \bibinfo{author}{\bibfnamefont{K.}~\bibnamefont{Burnett}},
  \bibinfo{journal}{J.\ Opt.\ B: Quant.\ S.\ Opt.}
  \textbf{\bibinfo{volume}{5}}, \bibinfo{pages}{S50} (\bibinfo{year}{2003});
\bibinfo{author}{\bibfnamefont{M.}~\bibnamefont{Rigol}} \bibnamefont{and}
  \bibinfo{author}{\bibfnamefont{A.}~\bibnamefont{Muramatsu}},
  \bibinfo{journal}{Phys.\ Rev.\ Lett.} \textbf{\bibinfo{volume}{93}},
  \bibinfo{pages}{230404} (\bibinfo{year}{2004});
\bibinfo{author}{\bibfnamefont{A.~M.} \bibnamefont{Rey}},
  \bibinfo{author}{\bibfnamefont{G.}~\bibnamefont{Pupillo}},
  \bibinfo{author}{\bibfnamefont{C.~W.} \bibnamefont{Clark}}, \bibnamefont{and}
  \bibinfo{author}{\bibfnamefont{C.~J.} \bibnamefont{Williams}},
  \bibinfo{journal}{Phys.\ Rev.\ A} \textbf{\bibinfo{volume}{72}},
  \bibinfo{pages}{033616} (\bibinfo{year}{2005});
\bibinfo{author}{\bibfnamefont{G.}~\bibnamefont{Mazzarella}},
  \bibinfo{author}{\bibfnamefont{S.~M.} \bibnamefont{Giampaolo}},
  \bibnamefont{and}
  \bibinfo{author}{\bibfnamefont{F.}~\bibnamefont{Illuminati}},
  \bibinfo{journal}{Phys.\ Rev.\ A} \textbf{\bibinfo{volume}{73}},
  \bibinfo{pages}{013625} (\bibinfo{year}{2006}).

\bibitem[{Gro({\natexlab{i}})}]{Group8}
\bibinfo{author}{\bibfnamefont{S.~R.} \bibnamefont{Clark}} \bibnamefont{and}
  \bibinfo{author}{\bibfnamefont{D.}~\bibnamefont{Jaksch}},
  \bibinfo{journal}{New.\ J.\ Phys.} \textbf{\bibinfo{volume}{8}},
  \bibinfo{pages}{160} (\bibinfo{year}{2006});
\bibinfo{author}{\bibfnamefont{M.}~\bibnamefont{Hild}},
  \bibinfo{author}{\bibfnamefont{F.}~\bibnamefont{Schmitt}}, \bibnamefont{and}
  \bibinfo{author}{\bibfnamefont{R.}~\bibnamefont{Roth}}, \bibinfo{journal}{J.\
  Phys.\ B} \textbf{\bibinfo{volume}{39}},
  \bibinfo{pages}{4547} (\bibinfo{year}{2006}).

\bibitem{Cederbaum1}
\bibinfo{author}{\bibfnamefont{O.~E.} \bibnamefont{Alon}},
  \bibinfo{author}{\bibfnamefont{A.~I.} \bibnamefont{Streltsov}} \bibnamefont{and}
  \bibinfo{author}{\bibfnamefont{L.~S.} \bibnamefont{Cederbaum}},
  \bibinfo{journal}{Phys.\ Rev.\ Lett.} \textbf{\bibinfo{volume}{95}},
  \bibinfo{pages}{030405} (\bibinfo{year}{2005});
  \bibinfo{journal}{Phys.\ Rev.\ Lett.} \textbf{\bibinfo{volume}{97}},
  \bibinfo{pages}{230403} (\bibinfo{year}{2006}).

\bibitem[{\citenamefont{{Gl\"uck} et~al.}(2002)\citenamefont{{Gl\"uck},
  Kolovsky, and Korsch}}]{GKK2002}
\bibinfo{author}{\bibfnamefont{M.}~\bibnamefont{{Gl\"uck}}},
  \bibinfo{author}{\bibfnamefont{A.~R.} \bibnamefont{Kolovsky}},
  \bibnamefont{and} \bibinfo{author}{\bibfnamefont{H.~J.}
  \bibnamefont{Korsch}}, \bibinfo{journal}{Phys.\ Rep.}
  \textbf{\bibinfo{volume}{366}}, \bibinfo{pages}{103} (\bibinfo{year}{2002}).

\bibitem[{\citenamefont{{M\"uller} et~al.}(2007)\citenamefont{{M\"uller},
  {F\"olling}, Widera, and Bloch}}]{MFWB2007}
\bibinfo{author}{\bibfnamefont{T.}~\bibnamefont{{M\"uller}}},
  \bibinfo{author}{\bibfnamefont{S.}~\bibnamefont{{F\"olling}}},
  \bibinfo{author}{\bibfnamefont{A.}~\bibnamefont{Widera}}, \bibnamefont{and}
  \bibinfo{author}{\bibfnamefont{I.}~\bibnamefont{Bloch}}
  (\bibinfo{year}{2007}), \eprint{cond-mat.other/0704.2856v1}.

\bibitem[{\citenamefont{Jaksch et~al.}(1998)\citenamefont{Jaksch, Bruder,
  Cirac, Gardiner, and Zoller}}]{JBCGZ1998}
\bibinfo{author}{\bibfnamefont{D.}~\bibnamefont{Jaksch}},
  \bibinfo{author}{\bibfnamefont{C.}~\bibnamefont{Bruder}},
  \bibinfo{author}{\bibfnamefont{J.~I.} \bibnamefont{Cirac}},
  \bibinfo{author}{\bibfnamefont{C.~W.} \bibnamefont{Gardiner}},
  \bibnamefont{and} \bibinfo{author}{\bibfnamefont{P.}~\bibnamefont{Zoller}},
  \bibinfo{journal}{Phys.\ Rev.\ Lett.} \textbf{\bibinfo{volume}{81}},
  \bibinfo{pages}{3108} (\bibinfo{year}{1998}).

\bibitem{Group20}
  \bibinfo{author}{\bibfnamefont{A.~R.} \bibnamefont{Kolovsky}} \bibnamefont{and}
  \bibinfo{author}{\bibfnamefont{A.}~\bibnamefont{Buchleitner}},
  \bibinfo{journal}{Europhys. Lett.} \textbf{\bibinfo{volume}{68}},
  \bibinfo{pages}{632} (\bibinfo{year}{2004});
  \bibinfo{author}{\bibfnamefont{A.~R.} \bibnamefont{Kolovsky}},
  \bibinfo{journal}{New\ J.\ Phys.} \textbf{\bibinfo{volume}{8}},
  \bibinfo{pages}{197} (\bibinfo{year}{2006}).

\bibitem[{\citenamefont{Haake}(2001)}]{Haake2001}
\bibinfo{author}{\bibfnamefont{F.}~\bibnamefont{Haake}},
  \emph{\bibinfo{title}{Quantum Signatures of Chaos}}
  (\bibinfo{publisher}{Springer}, \bibinfo{address}{Berlin},
  \bibinfo{year}{2001}).

\bibitem[{\citenamefont{Press et~al.}(1993)\citenamefont{Press, Teukolsky,
  Vetterling, and Flannery}}]{NumericalRecipes}
\bibinfo{author}{\bibfnamefont{W.~H.} \bibnamefont{Press}},
  \bibinfo{author}{\bibfnamefont{S.~A.} \bibnamefont{Teukolsky}},
  \bibinfo{author}{\bibfnamefont{W.~T.} \bibnamefont{Vetterling}},
  \bibnamefont{and} \bibinfo{author}{\bibfnamefont{B.~P.}
  \bibnamefont{Flannery}}, \emph{\bibinfo{title}{Numerical Recipes}}
  (\bibinfo{publisher}{Cambridge University Press},
  \bibinfo{address}{Cambridge}, \bibinfo{year}{1993}).

\bibitem[{\citenamefont{Prosen and Robnik}(1993)}]{PR1993}
\bibinfo{author}{\bibfnamefont{T.}~\bibnamefont{Prosen}} \bibnamefont{and}
  \bibinfo{author}{\bibfnamefont{M.}~\bibnamefont{Robnik}},
  \bibinfo{journal}{J.\ Phys.\ A} \textbf{\bibinfo{volume}{26}},
  \bibinfo{pages}{2371} (\bibinfo{year}{1993}).

\bibitem[{\citenamefont{Mehta}(1991)}]{Mehta1991}
\bibinfo{author}{\bibfnamefont{M.}~\bibnamefont{Mehta}},
  \emph{\bibinfo{title}{Random Matrices and the Statistical Theory of Energy
  Levels}} (\bibinfo{publisher}{Academic Press}, \bibinfo{address}{New York},
  \bibinfo{year}{1991}).

\bibitem[{Gro({\natexlab{k}})}]{Group12}
\bibinfo{author}{\bibfnamefont{T.}~\bibnamefont{Bergeman}},
  \bibinfo{author}{\bibfnamefont{M.~G.} \bibnamefont{Moore}}, \bibnamefont{and}
  \bibinfo{author}{\bibfnamefont{M.}~\bibnamefont{Olshanii}},
  \bibinfo{journal}{Phys.\ Rev.\ Lett.} \textbf{\bibinfo{volume}{91}},
  \bibinfo{pages}{163201} (\bibinfo{year}{2003}).
  
\bibitem{KMSGE2005}
\bibinfo{author}{\bibfnamefont{M.}~\bibnamefont{K\"ohl}},
  \bibinfo{author}{\bibfnamefont{H.}~\bibnamefont{Moritz}},
  \bibinfo{author}{\bibfnamefont{T.}~\bibnamefont{St\"oferle}},
  \bibinfo{author}{\bibfnamefont{K.}~\bibnamefont{G\"unter}}  \bibnamefont{and}
  \bibinfo{author}{\bibfnamefont{T.}~\bibnamefont{Esslinger}},
  \bibinfo{journal}{Phys.\ Rev.\ Lett.} \textbf{\bibinfo{volume}{94}},
  \bibinfo{pages}{080403} (\bibinfo{year}{2005}).

\bibitem[{Gro({\natexlab{l}})}]{Group10}
\bibinfo{author}{\bibfnamefont{S.}~\bibnamefont{Wimberger}},
  \bibinfo{author}{\bibfnamefont{R.}~\bibnamefont{Mannella}},
  \bibinfo{author}{\bibfnamefont{O.}~\bibnamefont{Morsch}},
  \bibinfo{author}{\bibfnamefont{E.}~\bibnamefont{Arimondo}},
  \bibinfo{author}{\bibfnamefont{A.~R.} \bibnamefont{Kolovsky}},
  \bibnamefont{and}
  \bibinfo{author}{\bibfnamefont{A.}~\bibnamefont{Buchleitner}},
  \bibinfo{journal}{Phys.\ Rev.\ A} \textbf{\bibinfo{volume}{72}},
  \bibinfo{pages}{063610} (\bibinfo{year}{2005});
\bibinfo{author}{\bibfnamefont{S.}~\bibnamefont{Wimberger}},
  \bibinfo{author}{\bibfnamefont{P.}~\bibnamefont{Schlagheck}},
  \bibnamefont{and} \bibinfo{author}{\bibfnamefont{R.}~\bibnamefont{Mannella}},
  \bibinfo{journal}{J.\ Phys.\ B}
  \textbf{\bibinfo{volume}{39}}, \bibinfo{pages}{729} (\bibinfo{year}{2006});
\bibinfo{author}{\bibfnamefont{D.}~\bibnamefont{Witthaut}},
  \bibinfo{author}{\bibfnamefont{E.~M.}~\bibnamefont{Graefe}}
  \bibnamefont{and} \bibinfo{author}{\bibfnamefont{H.~J.}~\bibnamefont{Korsch}},
  \bibinfo{journal}{Phys.\ Rev.\ A}
  \textbf{\bibinfo{volume}{73}}, \bibinfo{pages}{063609} (\bibinfo{year}{2006});  
\bibinfo{author}{\bibfnamefont{D.}~\bibnamefont{Witthaut}},
  \bibinfo{author}{\bibfnamefont{E.~M.}~\bibnamefont{Graefe}},
  \bibinfo{author}{\bibfnamefont{S.}~\bibnamefont{Wimberger}},
  \bibnamefont{and} \bibinfo{author}{\bibfnamefont{H.~J.}~\bibnamefont{Korsch}},
  {\it ibid.} 
  \textbf{\bibinfo{volume}{75}}, \bibinfo{pages}{013617} (\bibinfo{year}{2007}).

\bibitem[{\citenamefont{Kottos}(2005)}]{Kottos2005}
\bibinfo{author}{\bibfnamefont{T.}~\bibnamefont{Kottos}}, \bibinfo{journal}{J.\
  Phys.\ A} \textbf{\bibinfo{volume}{38}}, \bibinfo{pages}{10761}
  (\bibinfo{year}{2005}).

\bibitem{Group21}
\bibinfo{author}{\bibfnamefont{S.}~\bibnamefont{Wimberger}} \bibnamefont{and}
  \bibinfo{author}{\bibfnamefont{A.}~\bibnamefont{Buchleitner}},
  \bibinfo{journal}{J.\ Phys.\ A} \textbf{\bibinfo{volume}{34}},
  \bibinfo{pages}{7181} (\bibinfo{year}{2001});
\bibinfo{author}{\bibfnamefont{S.}~\bibnamefont{Wimberger}},
  \bibinfo{author}{\bibfnamefont{A.}~\bibnamefont{Krug}} \bibnamefont{and}
  \bibinfo{author}{\bibfnamefont{A.}~\bibnamefont{Buchleitner}},
  \bibinfo{journal}{Phys.\ Rev.\ Lett.} \textbf{\bibinfo{volume}{89}},
  \bibinfo{pages}{263601} (\bibinfo{year}{2002}).

\bibitem[{\citenamefont{Fyodorov and Sommers}(1996)}]{FS1996}
\bibinfo{author}{\bibfnamefont{Y.~V.} \bibnamefont{Fyodorov}} \bibnamefont{and}
  \bibinfo{author}{\bibfnamefont{H.}~\bibnamefont{Sommers}},
  \bibinfo{journal}{JETP Lett.} \textbf{\bibinfo{volume}{63}},
  \bibinfo{pages}{1026} (\bibinfo{year}{1996}).

\bibitem{Cederbaum2}
\bibinfo{author}{\bibfnamefont{A.~I.} \bibnamefont{Streltsov}}, 
\bibinfo{author}{\bibfnamefont{O.~E.} \bibnamefont{Alon}} \bibnamefont{and}
  \bibinfo{author}{\bibfnamefont{L.~S.} \bibnamefont{Cederbaum}},
  \bibinfo{journal}{Phys.\ Rev.\ Lett.} \textbf{\bibinfo{volume}{99}},
  \bibinfo{pages}{030402} (\bibinfo{year}{2007}).


\bibitem[{\citenamefont{Slater}(1952)}]{Slater1952}
\bibinfo{author}{\bibfnamefont{J.~C.} \bibnamefont{Slater}},
  \bibinfo{journal}{Phys.\ Rev.} \textbf{\bibinfo{volume}{87}},
  \bibinfo{pages}{807} (\bibinfo{year}{1952}).

\end{thebibliography}
\end{document}